# Accuracy and Efficiency Benchmarks of Pretrained Machine Learning Potentials for Molecular Simulations

Peter Eastman[1*], Evan Pretti[1], and Thomas E. Markland[1*]

[1]Department of Chemistry, Stanford University, Stanford, CA 94305, USA

*Corresponding authors: Peter Eastman (peastman@stanford.edu), Thomas E. Markland (tmarkland@stanford.edu)

# Abstract

The rapid development of pretrained Machine Learning Interatomic Potentials (MLIPs) that cover a wide range of molecular species has made it challenging to select the best model for a given application. We benchmark 15 pretrained MLIPs, evaluating each one on accuracy, speed, memory use, and ability to produce stable simulations. This provides an objective basis for practitioners to select the most appropriate MLIP for their own simulations, and offers insight into which factors most strongly influence model accuracy. We find that the number of model parameters and the size of the training set are both strongly correlated with accuracy, but observe no benefit from including explicit Coulomb energy terms. Speed and memory use are determined as much by the model architecture as by the size of the model.

# Introduction

Machine Learning Interatomic Potentials[1–3] (MLIPs) are a popular tool for molecular simulation. A machine learning model is trained to reproduce a large dataset of forces and energies computed with a high level quantum chemistry method. The trained model can often approach the accuracy of the original method while being orders of magnitude faster.[4]

In the past few years, several large datasets of quantum chemistry calculations on diverse molecules have become available.[5–8] This has enabled the creation of foundation models: pretrained models capable of simulating arbitrary molecules across large ranges of chemical space. The number of foundation models available to the community has exploded, offering powerful new tools for running simulations at higher accuracy than was possible before.[9–18]

The abundance of pretrained foundation models has created its own challenge: it can be difficult for a user to know which of the available models is the best choice for their application. The developers of each model typically publish benchmarks to demonstrate the accuracy and speed of their models, but there is no standardization to them. When two publications use different measures of accuracy on different test sets, it is impossible to tell which of the two models is actually more accurate.  Speed is usually measured by simulating different systems on different hardware, again making comparison impossible.

The more common test sets also have limitations that reduce their usefulness.  Many models are only benchmarked on neutral molecules[11,13], providing no information about how accurate they are likely to be on charged molecules.  They often contain only tiny molecules, providing no information about how accurate they are likely to be on larger molecules.  For example, the popular MD17 benchmark[19] only contains molecules with up to 21 atoms.  The GMTKN55 benchmark[20], which aims to provide very broad coverage of interaction types, contains no system with more than 60 atoms, and the vast majority of its test systems have less than 30 atoms.

Memory use is another critical aspect of MLIP evaluation.  The memory required to evaluate an MLIP can be orders of magnitude higher than for classical force fields, and they usually are computed on GPUs that may have quite limited memory.  Current generation desktop GPUs typically have between 12 and 24 GB of memory, while server GPUs may have several times that.  This sets an absolute upper limit on the largest system a particular MLIP can be applied to, yet it is rarely discussed or reported by model developers.

In this paper, we provide an objective, uniform evaluation of a large collection of MLIPs.  This provides guidance to practitioners looking to select a model for their own simulations, and illuminates the tradeoffs made by different models.  Our goal is not to select a single best model; different applications will have different requirements that, for example, prioritize speed versus accuracy in different ways.  We also do not aim to exhaustively investigate the strengths and weaknesses of each model.  We provide a few high level metrics that reveal major differences in model performance and allow them to be objectively compared.  Exploring the detailed reasons for the differences is beyond the scope of this work.

A few other benchmarks for pretrained MLIPs have been created that are conceptually similar to this one.  The FAIR Chemistry Leaderboard[21] provides several metrics to evaluate the accuracy of models.  It reports only accuracy, not speed or memory use.  Although it is intended to be an open benchmark, at the time of writing it contains almost exclusively models developed by a single organization (Meta).  MLIPAudit[22] similarly reports various metrics for accuracy, but not speed or memory use, and it is designed primarily for models implemented in one particular software package.  CatBench[23] includes a larger selection of models and reports both accuracy and speed (but not memory use), but it is targeted specifically at applications in computational catalysis.  All of these benchmarks are structured as online leaderboards, providing raw data but little analysis or commentary.

# Methods

## Evaluated MLIPs

The MLIPs included in this evaluation are listed in Table 1. They were selected based on the following criteria.

| Model | Elements | Parameters (million) | Training Samples (million) | Level of Theory | Charges? | Permissive License? |
|---|---|---|---|---|---|---|
| AceFF-1.1[9] | 12 | 0.5 | 11 | ωB97M-V/def2-TZVPPD | Yes | Yes |
| AceFF-2.0[9] | 12 | 1.0 | 12 | ωB97M-V/def2-TZVPPD | Yes | Yes |
| AIMNet2[10] | 14 | 2.2 | 20 | ωB97M-D3/def2-TZVPP | Yes | Yes |
| Egret-1[11] | 10 | 3.6 | 0.95 | ωB97M-D3BJ/def2-TZVPPD | No | Yes |
| FeNNix-Bio1(S)[12] | 12 | 7.4 | 2.2 | ωB97M-D3BJ/aug-cc-pVTZ | Yes | No |
| FeNNix-Bio1(M)[12] | 12 | 9.5 | 2.2 | ωB97M-D3BJ/aug-cc-pVTZ | Yes | No |
| MACE-MH-1[14] | 89 | 6.4 | 116 | ωB97M-D3BJ/def2-TZVPPD | No | No |
| MACE-OFF23(S)[13] | 10 | 0.7 | 0.95 | ωB97M-D3BJ/def2-TZVPPD | No | No |
| MACE-OFF23(L)[13] | 10 | 4.7 | 0.95 | ωB97M-D3BJ/def2-TZVPPD | No | No |
| MACE-OFF24(M)[13] | 10 | 1.4 | 1.16 | ωB97M-D3BJ/def2-TZVPPD | No | No |
| MACE-OMOL-0[24] | 89 | 52 | 100 | ωB97M-V/def2-TZVPD | Yes | No |
| MACELES-OFF[16] | 10 | 1.9 | 0.95 | ωB97M-D3BJ/def2-TZVPPD | No | Yes |
| Orb-v3-omol[17] | 89 | 26 | 100 | ωB97M-V/def2-TZVPD | Yes | Yes |
| UMA-s-1.1[18] | 89 | 150 | 484 | ωB97M-V/def2-TZVPD | Yes | Yes |
| UMA-m-1.1[18] | 89 | 1400 | 484 | ωB97M-V/def2-TZVPD | Yes | Yes |

Table 1. The MLIP models included in this study. Information listed for each one includes: the number of chemical elements it supports; the number of parameters it contains; the number of samples it was trained on; the level of theory it is trained to predict; whether it is designed to support charged systems; and whether it is distributed under a permissive license that allows commercial use.

- We only consider models that are suitable for molecular applications. We do not include ones designed primarily to simulate materials rather than molecules.
- They must support enough elements to be useful for a wide range of molecular simulations. All of the tested models support at least 10 elements.
- They must be able to conserve energy. Some models attempt to predict forces directly rather than calculating them as the gradient of energy. This can lead to faster performance[17], but it means a simulation does not sample a well defined potential energy surface. This makes them unsuitable for many applications.
- We do not require that they can simulate reactivity. Some models are trained to accurately reproduce transition states to bond formation, while others are trained only on whole stable molecules. Our evaluation only considers accuracy on stable molecules.

The MLIPs listed in Table 1 were trained to reproduce different levels of theory, but all of them are quite similar: the ωB97M DFT functional[25] combined with a triple-zeta basis set. They vary in the choice of dispersion correction[26–28] and in the precise basis set used. We find that these differences have little effect on the overall accuracy of the models. The MACE-MH-1, UMA-s-1.1, and UMA-m-1.1 models were trained on multiple datasets generated at different levels of theory. These models allow the user to select which level of theory the output should correspond to. We have selected the one most similar to our test set, as shown in Table 1.

Some models were trained only on neutral molecules. Others included charged molecules in their training sets, and allow the user to specify the total charge when performing calculations. Table 1 lists whether each model is designed to support charged molecules (it was trained on charged molecules and allows the specification of the charge of the system). Models that were trained only on neutral molecules can still be used to compute energies for charged molecules, but it is plausible that their accuracy will be lower in that case than models designed to support charges.

An unfortunate reality is that users must consider not only the technical features of each model, but also legal requirements. Many MLIPs are provided under restrictive licenses that forbid commercial use. This is shown in Table 1, since for many users it rules out these models regardless of their technical merits.

## Accuracy Evaluation

To compare the accuracy of the different MLIPs we used the SPICE test set.[6,29] This is a collection of 800 molecules and dimers with 10 conformations for each one, evaluated at the ωB97M-D3BJ/def2-TZVPPD level of theory. They were not used in training any of the models. It is composed of four subsets.

- 200 small ligands drawn from Ligand Expo with between 40 and 50 atoms.
- 200 large ligands drawn from Ligand Expo with between 70 and 80 atoms.
- 200 pentapeptides with randomly chosen sequences. They have between 68 and 110 atoms.
- 200 dimers derived from Protein Data Bank structures. Each one consists of a ligand and a single amino acid with which it interacts. They have between 34 and 72 atoms.

Of the 800 molecules and dimers, 617 are neutral and the other 183 are charged. Charges vary between -4 and +2, but the vast majority are between -1 and +1.

A single molecule contains boron, which is not supported by some of the tested MLIPs. That molecule was omitted when testing those models.

Because of the differences in level of theory and the arbitrary zero point of energy, we do not expect MLIPs to reproduce absolute energies. Instead we focus on energy differences between conformations of the same molecule, which should be much less sensitive to those factors. For each molecule or dimer, we compute the mean absolute error (MAE) in energy differences as

$$MAE = \frac{1}{N(N-1)/2} \sum_{i<j} \left| (E_i - E_j) - (E_i^R - E_j^R) \right| \tag{1}$$

where N is the number of conformations (10 for all but a handful of molecules for which one DFT calculation failed to converge), $E_i$ is the energy of conformation i predicted by the MLIP, and $E^R_i$ is the reference energy for conformation i found in the test set.

We also tried measuring errors in forces, but we found the differences in what level of theory each model was trained on led to small but still measurable differences. A clear separation emerged between models whose training sets used exactly the same level of theory as the test set versus ones that used a similar but not identical level of theory. No such separation occurs when computing errors in energy differences, which appear to be less sensitive to these differences.

For purposes of this study, we treat the DFT outputs as exact and judge MLIPs on their ability to reproduce them. This is justified by the fact that all models were trained to reproduce DFT data at similar levels of theory, and the best result we can hope for is that they reproduce that level of theory.

We do not attempt to compare to experimental results such as measured free energies. That would be a much more complicated comparison, depending not only on the intrinsic accuracies of the models but also on the detailed simulation protocol used to compute the results, the lengths of the simulations, the inclusion of nuclear quantum effects, etc. It would introduce many sources of error that would be common to all models and therefore not relevant to comparing them. Ultimately, the only meaningful difference in model accuracy is their ability to reproduce the DFT functional they were trained on.

## Speed and Memory Evaluation

To evaluate the speed and memory use of each MLIP, we used the Atomic Simulation Environment[30] (ASE) to run a series of short simulations on an NVIDIA H100 GPU with 80 GB of memory. The time required to complete the simulation was recorded, giving a speed in steps/second. In addition, the amount of GPU memory in use was recorded just before creating the model, and again at the end of the simulation. The difference between the two was recorded as the required memory use.

To measure performance on small molecules, we simulated three molecules from the test set: one with 50 atoms (PDB chemical component ID WP0), one with 75 atoms (PDB chemical component ID YLY), and one with 100 atoms (the peptide with sequence PRO-THR-ILE-TYR-PHE).

To measure performance on larger systems, we simulated a series of water boxes whose widths varied from 2 to 6 nm. The total number of atoms in these boxes varied from 774 to 21,384. Most models were unable to simulate the largest water boxes due to running out of memory.

Each system was first energy minimized with L-BFGS[31] to remove any large forces, then equilibrated for 100 steps to let the model perform any necessary initialization. We then recorded the time to integrate 400 steps for the molecules or 100 steps for the water boxes.

Further details of the simulations are found in the Appendix.

## Simulation Stability

Another requirement for an MLIP to be useful is that it can produce stable simulations. A variety of problems with a potential energy surface can lead to instability, for example excessively large or rapidly changing forces that produce numerical errors, or unnaturally low barriers that allow bonds to break.

To screen for problems in simulation stability, we simulated the same 50 atom molecule as before (WP0), but solvated it with 212 water molecules to give a total of 686 atoms. For each MLIP, we began with L-BFGS energy minimization. We then set the velocities to random values drawn from a Boltzmann distribution at 400K, equilibrated it for 10 ps, and finally simulated it for 100 ps (100,000 steps) using a Langevin integrator at 400K with a 1 fs time step and 1 $ps^{-1}$ friction coefficient. The elevated temperature was chosen to amplify any stability issues and increase the chance of them being apparent in a short simulation.

The instantaneous temperature was computed after each step, and the mean and maximum instantaneous temperature observed during the simulation was recorded. Numerical instability can lead to temperature spikes or, in more extreme cases, a steady increase in temperature. The observation of excessively high temperature at any point in the simulation is therefore an indicator of possible instability.

At the end of the simulation, all covalent bonds in the system were checked to make sure they had not broken. Specifically, we checked that the length of each bond had not increased by more than 0.5 Å.

# Results

Results for accuracy are shown in Table 2 and Figure 1. Models are sorted by their MAE averaged over the whole test set. The MAE for each of the subsets that make up the test set is also shown. Finally, the test set is divided into neutral and charged systems, and the average MAE is shown for each.

| Model | Overall | Small Lig. | Large Lig. | Peptides | Dimers | Neutral | Charged |
|---|---|---|---|---|---|---|---|
| UMA-m-1.1 | 0.53 | 0.44 | 0.65 | 0.66 | 0.39 | 0.52 | 0.59 |
| UMA-s-1.1 | 0.61 | 0.51 | 0.7 | 0.67 | 0.57 | 0.57 | 0.76 |
| Orb-v3-omol | 0.7 | 0.58 | 0.84 | 0.74 | 0.66 | 0.65 | 0.88 |
| MACE-OMOL-0 | 0.9 | 0.59 | 0.94 | 1.09 | 0.98 | 0.79 | 1.25 |
| MACE-MH-1 | 1.5 | 1.02 | 1.57 | 1.5 | 1.9 | 1.16 | 2.64 |
| Egret-1 | 1.7 | 1.02 | 1.68 | 1.53 | 2.56 | 1.19 | 3.41 |
| MACE-OFF23(L) | 1.73 | 1.06 | 1.73 | 1.73 | 2.4 | 1.24 | 3.41 |
| FeNNix-Bio1(M) | 1.88 | 1.56 | 1.95 | 1.5 | 2.53 | 1.54 | 3.06 |
| MACE-OFF24(M) | 1.95 | 1.35 | 1.97 | 1.99 | 2.48 | 1.47 | 3.56 |
| MACELES-OFF | 1.96 | 1.17 | 2.52 | 1.66 | 2.49 | 1.52 | 3.45 |
| FeNNix-Bio1(S) | 2.11 | 1.75 | 2.25 | 1.65 | 2.79 | 1.72 | 3.44 |
| AIMNet2 | 2.55 | 1.88 | 2.95 | 2.97 | 2.43 | 2.37 | 3.16 |
| AceFF-2.0 | 2.86 | 1.68 | 3.8 | 3.35 | 2.59 | 2.69 | 3.43 |
| AceFF-1.1 | 3.64 | 2.07 | 3.33 | 4.33 | 4.81 | 2.62 | 7.07 |
| MACE-OFF23(S) | 3.73 | 1.73 | 7.74 | 2.12 | 3.3 | 3.67 | 3.92 |

Table 2. Mean absolute error in kcal/mol. “Overall” is the mean error on the complete test set. Other columns correspond to subsets of the test set. Models are sorted by overall error. Cells are colored based on the range of values within each column, with yellow denoting the smallest errors and blue the largest.

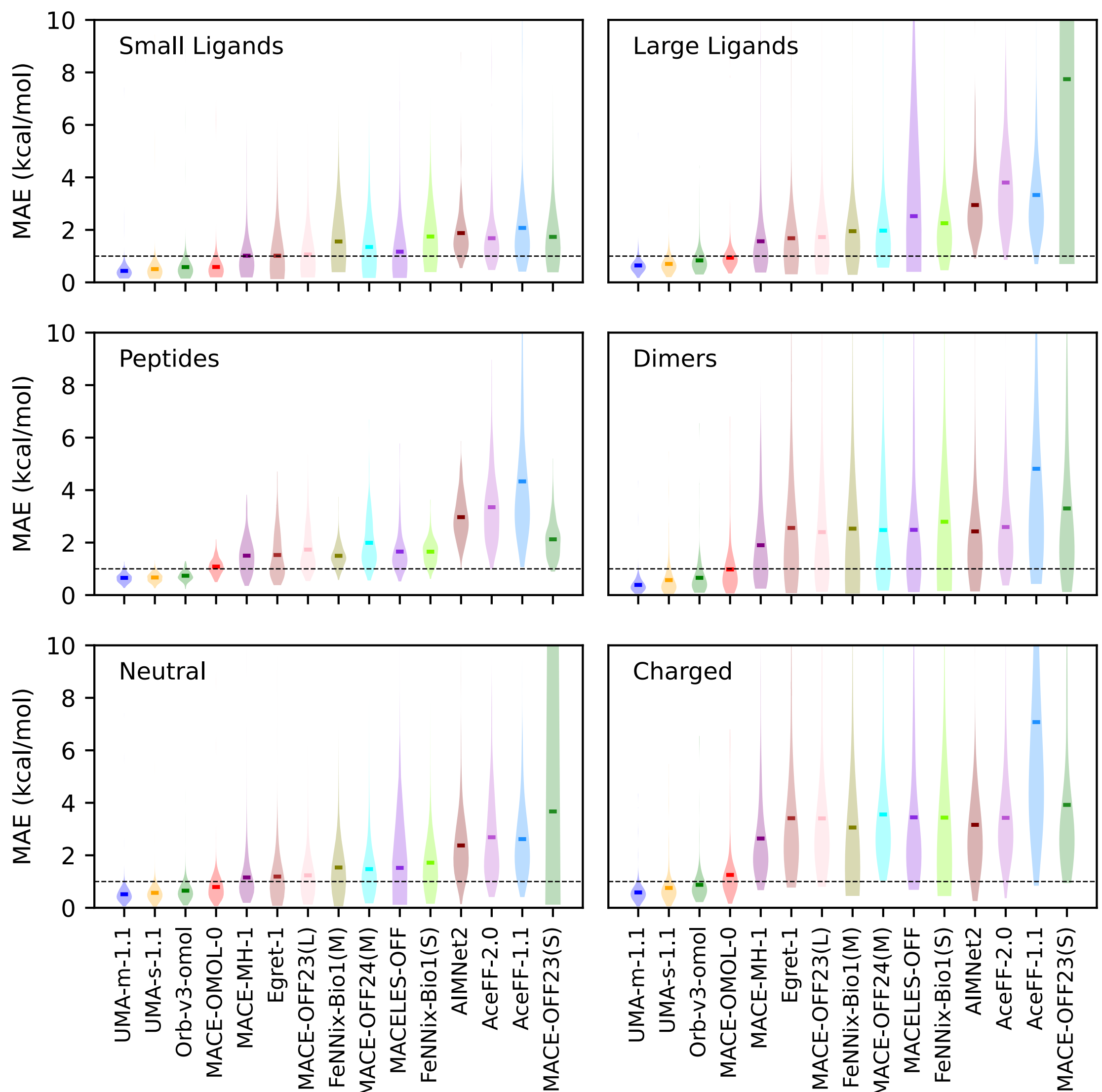

Figure 1. Distribution of errors for each model on each subset. The horizontal bars indicate the mean error over all molecules in the subset. The dashed line is at 1 kcal/mol, a commonly used standard for chemical accuracy. Models are sorted by overall error as shown in Table 2.

Most of the columns are strongly correlated with each other: a model that does well on one subset tends to do well on the others as well. Nonetheless, there are variations that are worth examining further, since they reveal specific strengths and weaknesses of particular models. Table 3 lists ratios of errors for significant pairs of subsets: large ligands to small ligands, and charged to neutral molecules.

| Model | Large/Small | Charged/Neutral |
|---:|---:|---:|
| AceFF-1.1 | 1.60 | 2.70 |
| AceFF-2.0 | 2.26 | 1.28 |
| AIMNet2 | 1.57 | 1.33 |
| Egret-1 | 1.66 | 2.87 |
| FeNNix-Bio1(M) | 1.25 | 1.99 |
| FeNNix-Bio1(S) | 1.29 | 2.00 |
| MACE-MH-1 | 1.54 | 2.28 |
| MACE-OFF23(L) | 1.63 | 2.75 |
| MACE-OFF23(S) | 4.47 | 1.07 |
| MACE-OFF24(M) | 1.46 | 2.41 |
| MACE-OMOL-0 | 1.61 | 1.58 |
| MACELES-OFF | 2.16 | 2.27 |
| Orb-v3-omol | 1.44 | 1.34 |
| UMA-m-1.1 | 1.48 | 1.13 |
| UMA-s-1.1 | 1.39 | 1.33 |

Table 3. Ratios of errors on subsets of the test set.

The ratio of large ligands to small ligands reveals how error increases with system size. On average, the large ligands are about 70% larger than the small ligands. We naively expect errors to be larger by an amount between 1.3 (corresponding to uncorrelated errors on atoms, so the total error scales as the square root of the number of atoms) and 1.7 (corresponding to correlated errors on atoms, so the total error scales as the number of atoms). Indeed, most ratios are within this range, but three models stand out. MACELES-OFF and AceFF-2.0 have faster than expected increases in error with ratios of 2.16 and 2.26, while MACE-OFF23(S) has dramatically higher error on the larger molecules with a ratio of 4.49.

This highlights the importance of testing models on a variety of molecule sizes. If a model has good accuracy on small molecules, that does not guarantee it will also be accurate on ones that are even modestly larger.

All models have higher errors on charged than neutral molecules. One might reasonably expect the models that were trained on charged molecules to be less affected by the presence of charges in the test set. This clearly is true: eight of the nine models trained on charged molecules (see Table 1) have a charged/neutral ratio of 2.0 or less, while five of the six models trained only on neutral molecules have ratios higher than that. Even so, the effect is weaker than might be expected. For example, AceFF-1.1 has one of the highest charged/neutral ratios despite being trained on charged molecules. The two FeNNix-Bio1 models likewise have relatively large ratios despite being trained on charged molecules. At the other extreme, the model with the very lowest charged/neutral ratio, MACE-OFF23(S), was trained only on neutral molecules.

Some MLIPs try to explicitly model the Coulomb interaction by incorporating an energy term that scales with distance as 1/r. This is intended both to improve accuracy on charged molecules, and also to improve scaling to large systems for which long range interactions are important. Five of the models we tested include such a term: AceFF-2.0, AIMNet2, FeNNix-Bio1(S), FeNNix-Bio1(M), and MACELES-OFF.

Based on this limited data, we see no clear evidence of either of these benefits. AceFF-2.0 and AIMNet2 do have particularly low charged/neutral ratios, but the FeNNix-Bio1 models have the highest ratios of any models that included charged systems in their training sets. For scaling with system size the situation is reversed: the FeNNix-Bio1 models have the lowest large/small ratios of any models, but AceFF-2.0 and MACELES-OFF have among the highest ratios. The presence of a 1/r term does not seem to be correlated with either of these measures.

It is possible a benefit would emerge when scaling to even larger systems, but if so, 80 atoms does not appear to be enough. The peptides are even larger with up to 110 atoms, and in that case too none of these five models stands out as particularly accurate. It is also notable that none of the seven models with the lowest overall error includes this term.

Comparing MACELES-OFF to MACE-OFF24(M) is especially revealing. Both models use the MACE architecture, they have similar numbers of parameters (MACELES-OFF is slightly larger), and they were trained on similar datasets. The main difference between them is the addition of a 1/r term in MACELES-OFF. Despite that difference, the overall accuracies of the two models are nearly identical.

Figure 2 compares accuracy to the number of parameters and number of training samples for each model. The correlation is very strong in both cases, especially the number of parameters which has an almost linear relationship to the model error. Despite the variety of architectures, datasets, and training methods represented by these models, the clear pattern is that larger models trained on larger datasets tend to be more accurate.

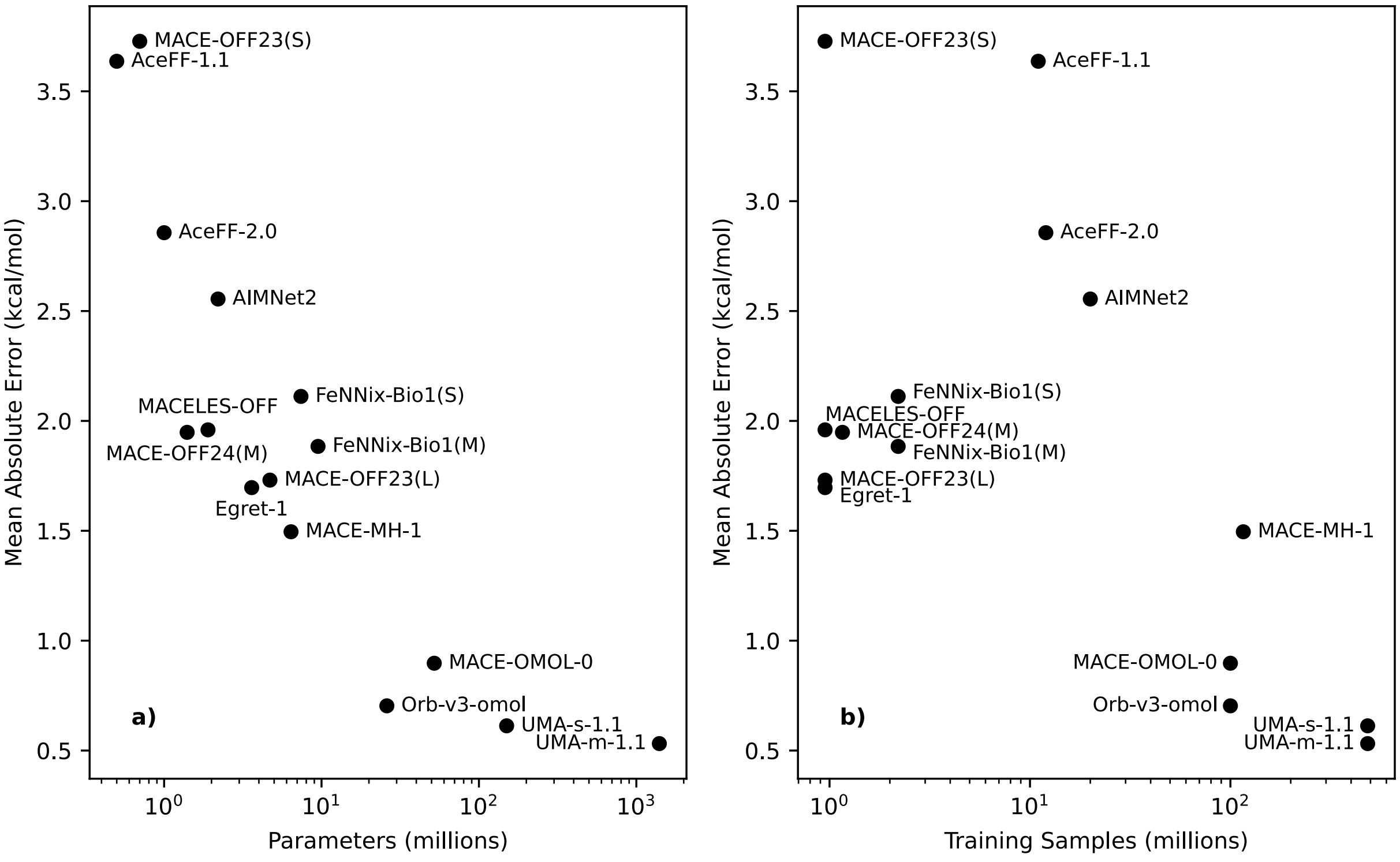

Figure 2. Model error versus a) number of parameters and b) number of training samples.

That is not to say that the architecture makes no difference. There is some notable variation around the average pattern. For example, FeNNix-Bio1(S) is less accurate than MACE-OFF24(M) and MACELES-OFF, despite having more parameters and a larger training set. Likewise, FeNNix-Bio1(M) is less accurate than Egret-1 and MACE-OFF23(L), despite having more parameters and a larger training set. It appears the MACE architecture is more parameter efficient and more data efficient than the FeNNix architecture. That does not necessarily mean it is better. As shown below, the FeNNix-Bio1 models are in fact faster and use less memory than the corresponding MACE models, despite having more parameters. Parameter counts and training set sizes are mainly of interest to model developers, not to users.

Data on speed and memory use is shown in Figures 3 and 4, and in Table 4. The speeds shown are the average over three replicate simulations. The error bars in Figure 3 correspond to the standard deviation of the three replicates. We do not show error bars for memory because the variation between runs is negligible.

We naively expect speed to decrease and memory use to increase in proportion to the number of atoms. For the most part this is what we observe, especially for the larger systems, but there are some notable exceptions.

UMA models offer a "turbo" mode that greatly improves speed, but also greatly increases memory use. We used turbo mode for these models wherever possible, but switched to the slower default mode when necessary to fit within the GPU's 80 GB limit. This accounts for the abrupt drop in speed and memory for these models. On a GPU with less memory, the drop would need to happen at a lower number of atoms.

Both of the FeNNix-Bio1 models have irregular performance on small molecules. Whereas most models have little variation in speed between runs, FeNNix-Bio1 models have enormous variation with speed on successive runs sometimes differing by as much as a factor of five. We hypothesized that this reflected insufficient equilibration to let the model fully initialize itself. We tried increasing the length of the equilibration to 1500 steps and the length of the simulation to 1000 steps, but neither of these eliminated the fluctuations.

The calculation time for most models appears to scale as O(N), as shown by the curves being parallel at the right side of Figure 3. The FeNNix-Bio1 models, however, have a steeper slope that indicates $O(N^2)$ scaling. AIMNet2 is still showing sub-linear scaling, suggesting it might become the very fastest model on even larger systems.

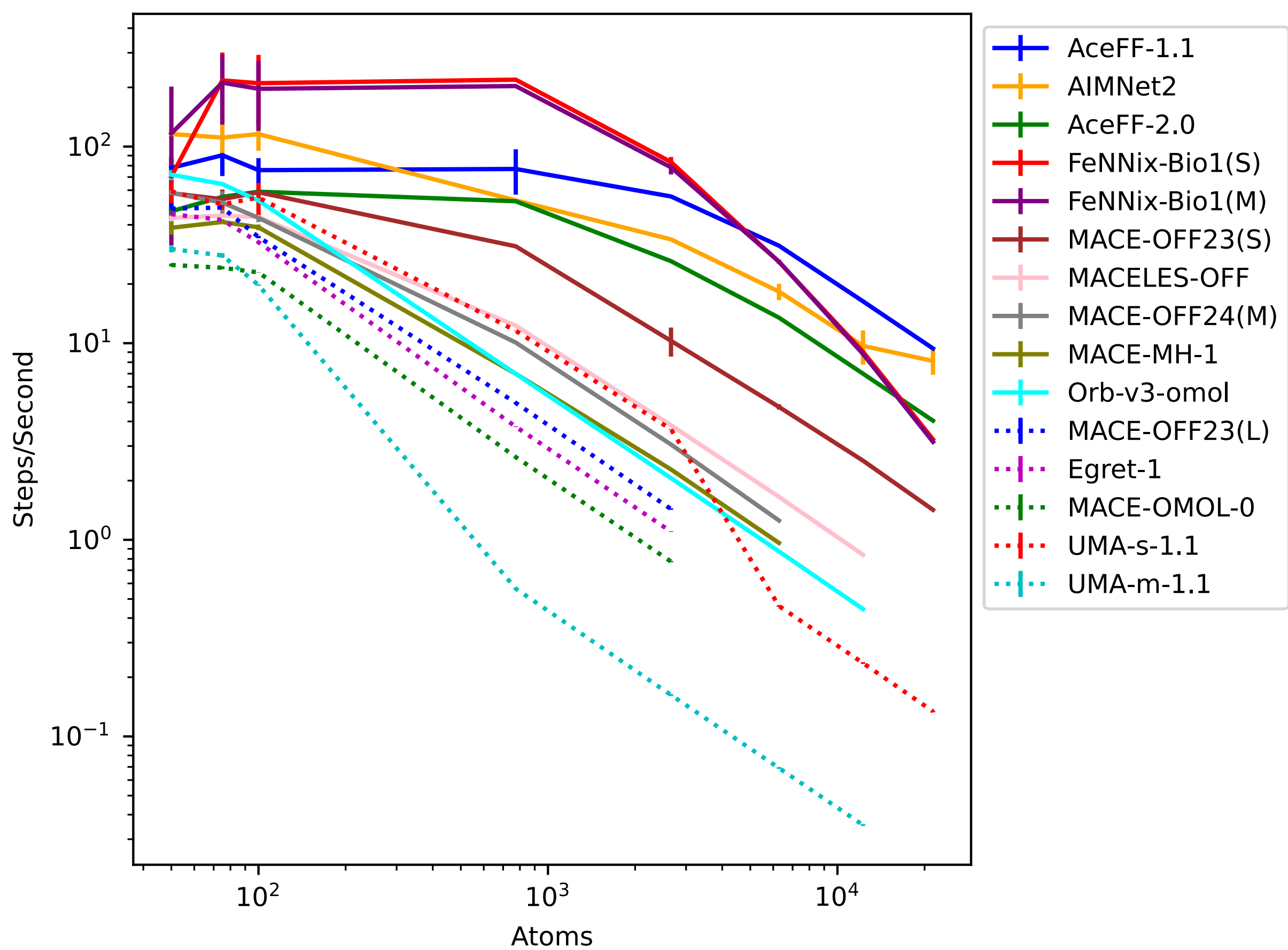

Figure 3. Speed versus number of atoms. The error bars indicate the standard deviation of three replicates.

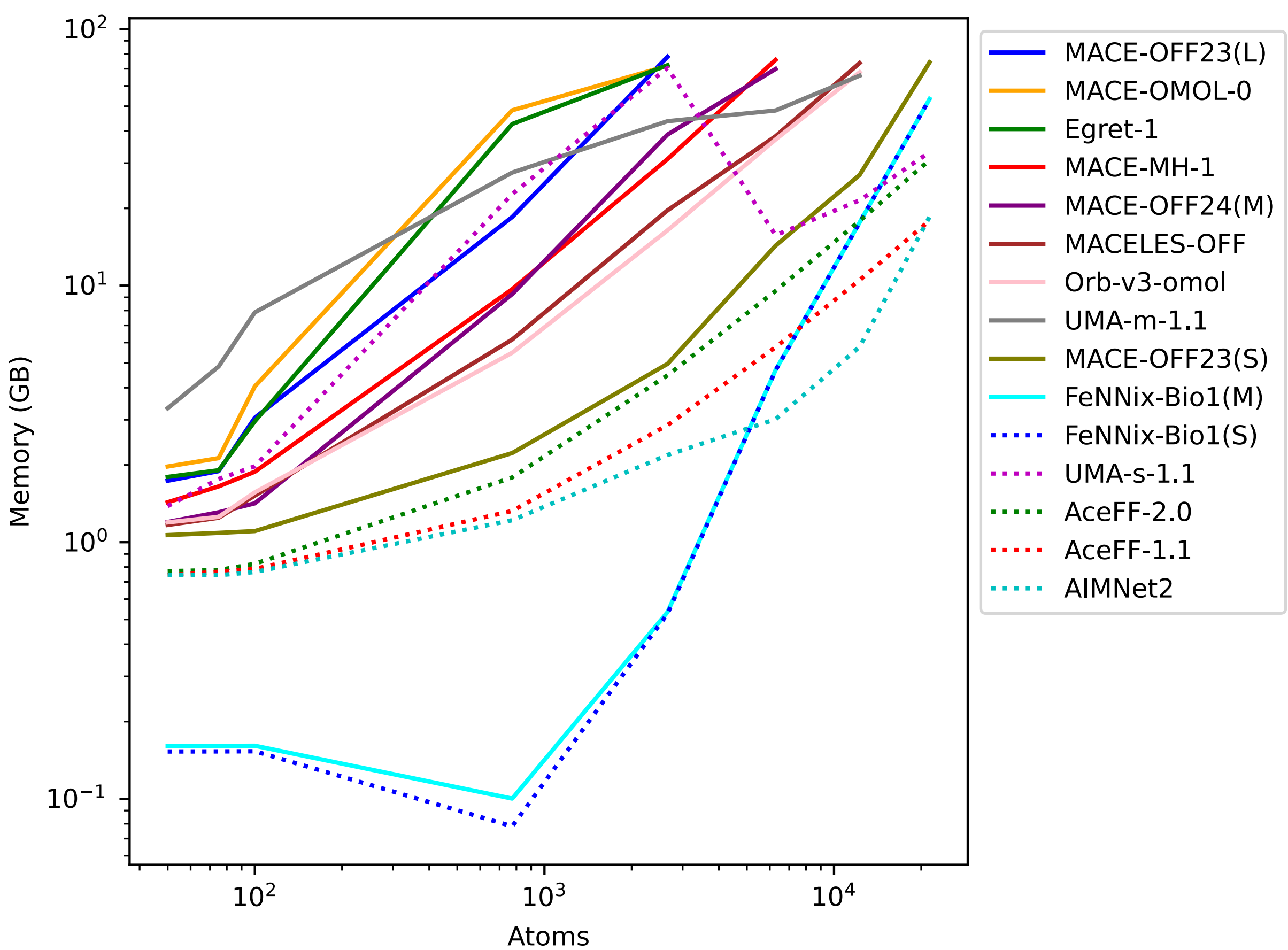

Figure 4. Memory required versus number of atoms.

| | **50** | **75** | **100** | **774** | **2661** | **6282** | **12,255** | **21,384** |
|---:|---:|---:|---:|---:|---:|---:|---:|---:|
| **AIMNet2** | 115.83 | 111.15 | 115.79 | 53.26 | 33.73 | 18.30 | 9.69 | 8.12 |
| **AceFF-1.1** | 77.99 | 90.36 | 75.89 | 76.93 | 55.78 | 31.32 | 16.33 | 9.40 |
| **AceFF-2.0** | 46.89 | 55.50 | 58.95 | 52.73 | 26.14 | 13.49 | 7.00 | 4.04 |
| **Egret-1** | 45.32 | 42.33 | 32.62 | 3.76 | 1.10 | - | - | - |
| **FeNNix-Bio1(M)** | 116.67 | 211.37 | 196.73 | 203.15 | 78.41 | 25.99 | 8.82 | 3.15 |
| **FeNNix-Bio1(S)** | 70.51 | 217.05 | 210.13 | 218.95 | 83.46 | 25.91 | 9.06 | 3.24 |
| **MACE-MH-1** | 38.65 | 41.36 | 38.85 | 6.99 | 2.28 | 0.96 | - | - |
| **MACE-OFF23(L)** | 48.35 | 48.95 | 34.68 | 4.97 | 1.43 | - | - | - |
| **MACE-OFF23(S)** | 57.94 | 53.99 | 58.58 | 31.10 | 10.28 | 4.74 | 2.53 | 1.42 |
| **MACE-OFF24(M)** | 58.19 | 51.91 | 43.40 | 10.08 | 3.05 | 1.25 | - | - |
| **MACE-OMOL-0** | 25.00 | 24.18 | 22.92 | 2.64 | 0.77 | - | - | - |
| **MACELES-OFF** | 43.44 | 44.54 | 43.87 | 12.27 | 3.83 | 1.65 | 0.84 | - |
| **Orb-v3-omol** | 71.98 | 64.46 | 53.16 | 6.99 | 2.06 | 0.87 | 0.44 | - |
| **UMA-m-1.1** | 30.14 | 27.94 | 19.77 | 0.56 | 0.16 | 0.07 | 0.04 | - |
| **UMA-s-1.1** | 59.15 | 50.85 | 55.05 | 11.57 | 3.65 | 0.46 | 0.24 | 0.13 |

Table 4. Speed in steps/second. Column headers are the number of atoms in each system. The simulations used a 1 fs step size, so 1 step/second corresponds to 0.0864 ns/day.

It is worth noting that memory use only depends very weakly on the size of a model. For example, UMA-s-1.1 is the second largest model with 150 million parameters, yet it can successfully simulate the largest water box without running out of memory. At the other extreme, Egret-1 with only 3.6 million parameters can only simulate the two smallest water boxes.

An important analysis is to compare speed to accuracy. This is one of the most essential tradeoffs when designing or selecting an MLIP. It is shown in Figure 5 for the 50 atom molecule and the 2661 and 12,255 atom water boxes. There is a clear correlation between them, but also a great deal of variation. The rule appears to be that higher accuracy requires a slower simulation, but slow models are not always more accurate.

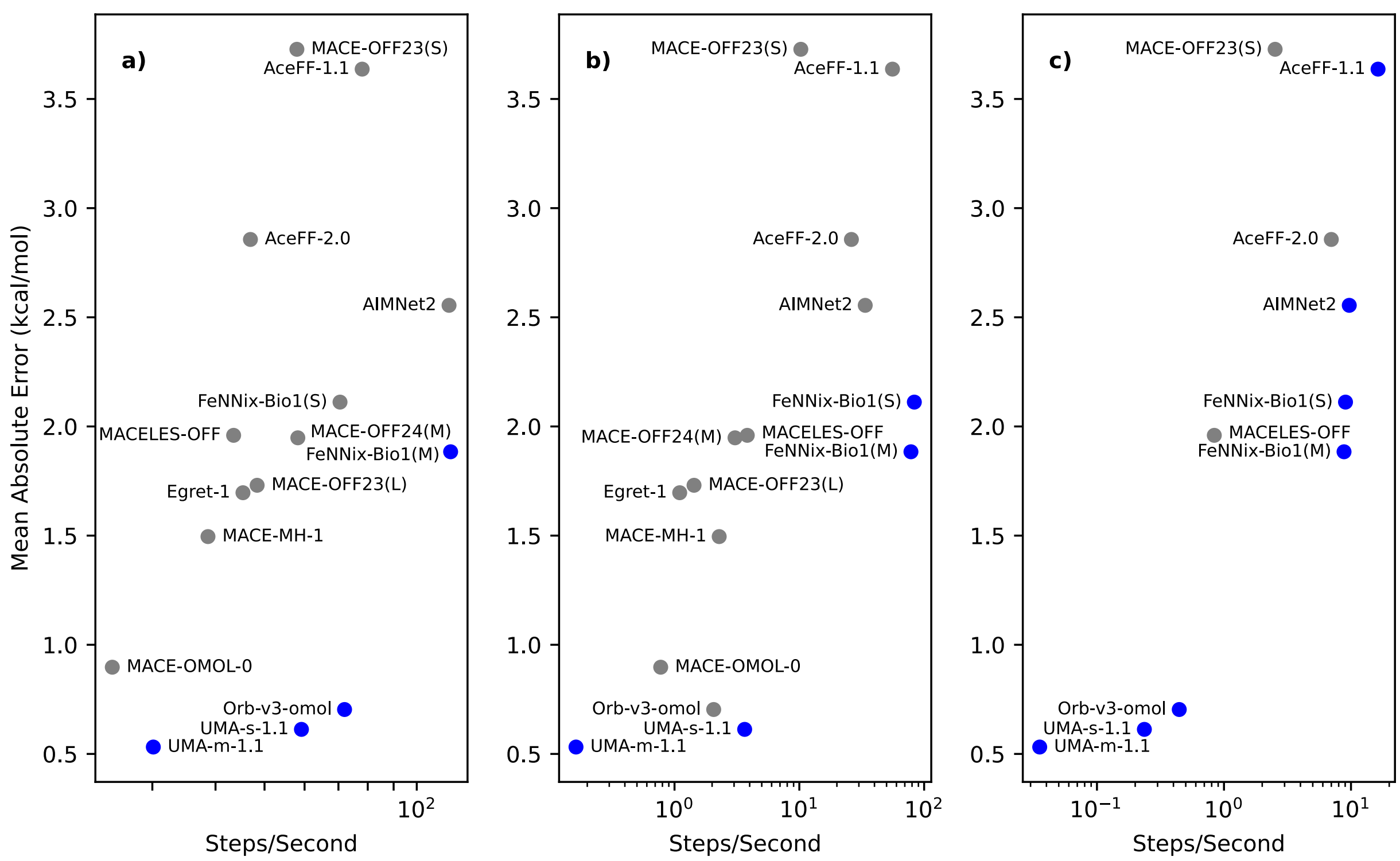

Figure 5. Model error on the test set versus speed on a) a 50 atom molecule, b) a 2661 atom water box, and c) a 12,255 atom water box. Models are shown in blue if no other single model is both faster and more accurate. These are the ones with the fastest performance for a given level of error.

In the simulations to test stability, no broken bonds were observed with any model. The average temperatures over the simulations varied from 391K to 397K. That the average temperature was always below the target temperature of 400K is expected, since the simulations began from an energy minimized structure and only had a very short (10 ps) equilibration. It is possible that stability problems might still appear in a longer simulation, or with a different test system. Based on these results, however, we see no indication of instability with any of these models.

# Discussion

MLIPs are a quickly changing field. This study is a snapshot of the state of the art at one moment in time. It aims to provide useful advice to practitioners looking to choose a model for their simulations, and also to guide the development of future MLIPs. For example, it may suggest which architectures have the best balance of speed and accuracy, or reveal weaknesses in specific models that can be addressed with further development.

Our goal is not to pick a single best model, but some models clearly stand out as more useful than others. Based on the results above, we can make some concrete recommendations.

Three models can reasonably be described as having "chemical accuracy", which we define as having an MAE $<$ 1 kcal/mol on every subset: UMA-m-1.1, UMA-s-1.1, and Orb-v3-omol. When sufficient GPU memory is available to use turbo mode, UMA-s-1.1 and Orb-v3-omol tend to have similar speed, with Orb-v3-omol being slightly faster on the small molecules and UMA-s-1.1 being faster on the water boxes. For systems that require too much memory to use turbo mode, UMA-s-1.1 becomes much slower and Orb-v3-omol is probably a better choice.

UMA-m-1.1 is the very most accurate model of all the ones tested, but the small difference in accuracy (0.17 kcal/mol in overall MAE compared to Orb-v3-omol) is unlikely to be very significant in most applications, and it comes at a high cost in speed. On all of the water boxes it is more than 12 times slower than Orb-v3-omol.

The four fastest models are FeNNix-Bio1(S), FeNNix-Bio1(M), AIMNet2, and AceFF-1.1, with their relative speeds varying based on the system being simulated. Of the four, AceFF-1.1 is significantly less accurate than the others while usually not being significantly faster, and therefore is difficult to recommend. The other three are all reasonable choices when speed is important and somewhat lower accuracy is acceptable.

We now offer advice for the development of future MLIPs. An unsurprising conclusion is that larger models trained on larger datasets tend to be more accurate. This must be balanced against the need for speed: all else being equal, larger models are more expensive to evaluate. An essential metric by which future models should be judged is their ratio of speed to accuracy. Sufficiently expensive models can already achieve excellent accuracy. Future progress should be defined as achieving similar accuracy at a lower computational cost.

There is no similar tradeoff in the choice of training set. Increasing the size of the training set is likely to be beneficial even without a corresponding increase in the model size, and does not make it any more expensive to evaluate. Given the availability of enormous datasets like OMol25[8] and OMat24[7], lack of training data should no longer be an issue for any model developer.

Charged molecules are very common in biology and chemistry; an MLIP that cannot model them accurately has limited value. Three techniques are commonly used to improve accuracy on them: 1) including charged systems in the training set, 2) providing a mechanism to specify the total charge of a system, and 3) including an energy term that scales with the distance between atoms as 1/r. The first two of these techniques appear to be clearly beneficial. On the other hand, we cannot identify any clear benefit from a 1/r term, either for simulating charged systems or for scaling to large molecules. That statement should not be interpreted as saying no benefit exists, only that there is not sufficient evidence to conclude that it does from the testing we have performed here. This question should be studied further.

Even MLIPs that use none of these techniques can still sometimes have acceptable accuracy on charged molecules. An example is MACE-MH-1, which achieves an MAE on charged molecules of 2.64 kcal/mol. This is much higher than its error on neutral molecules, but it is still low enough to be useful for many applications. That is remarkable given that it was trained exclusively on neutral systems, and it suggests that models can learn to generalize from uncharged to charged systems.

All of the benchmarks in this work involve isolated systems modeled entirely with MLIPs. In the future, an important application is likely to be ML/MM simulations in which a small piece of a system is modeled with an MLIP, while the rest is modeled with a conventional force field. It remains to be seen whether accuracy on isolated systems will translate to similar accuracy for ML/MM simulations. The answer will likely depend not only on the accuracy of the ML and MM models when used in isolation, but also on how compatible they are and on how they interact with each other. The interaction between the regions can be computed with a mechanical embedding in which their energies are computed independently; an electrostatic embedding in which the MM region polarizes the ML region; or a polarizable embedding in which the ML region also polarizes the MM region. The development of these methods and the study of their accuracy will be an important subject for future work.

# Acknowledgements

Research reported in this publication was supported by the National Institute of General Medical Sciences of the National Institutes of Health under award number 2R01GM140090 (PE and TEM); and the Wellcome Trust grant number 313301/Z/24/Z (EP). We thank Chris Iacovella and João Morado for helpful comments and suggestions. We thank Bingqing Cheng for help in identifying some unphysical conformations in the test set.

# Appendix: Simulation Details

Wherever possible, we attempted to use identical software versions for all benchmarks. Unfortunately, the models tested in this work are implemented in a variety of software packages that often have incompatible version requirements. The use of several different environments to run simulations was therefore unavoidable.

All of the models based on the MACE architecture used Python 3.12.12, CUDA 12.8, PyTorch 2.9.1, MACE 0.3.15, ASE 3.26.0, and cuEquivariance 0.8.0.

The UMA models used Python 3.12.12, CUDA 12.8, PyTorch 2.8.0, fairchem-core 2.19.0, and ASE 3.26.0.

The AceFF models used Python 3.12.12, CUDA 12.8, PyTorch 2.9.1, Triton 3.5.1, TorchMD-Net 3.0.3, and ASE 3.27.0.

AIMNet2 used Python 3.12.12, CUDA 12.8, PyTorch 2.9.1, aimnet 0.1.1, and ASE 3.22.1.

Orb-v3-omol used Python 3.12.12, CUDA 12.8, PyTorch 2.9.1, ASE 3.26.0, orb-models 0.6.2, and cuML 25.2.1.

The FeNNix-Bio1 models used Python 3.12.12, CUDA 12.8, JAX 0.8.1, FeNNol 2026.4.2, and ASE 3.26.0.

All calculations were performed in FP32 precision mode. We avoided the use of low precision modes such as TF32, because it is insufficiently accurate for many applications.

The scripts used to run the calculations described in this paper are available at https://github.com/peastman/mlipbenchmarks.